\documentclass[a4paper,11pt]{article}

\usepackage{amsmath}
\usepackage{amssymb}
\newcommand{\be}{\begin{equation}}
\newcommand{\ee}{\end{equation}}
\newcommand{\bea}{\begin{eqnarray}}
\newcommand{\eea}{\end{eqnarray}}

\def\ve{\varepsilon}
\def\Z{Z\hspace{-2mm}Z}
\def\tg{\mbox{tg}}

\def\ve{\varepsilon}

\def\ga{\gamma}

\def\nvec#1{\mbox{\bf #1}}

\def\vr{\nvec{r}}

\def\vv{\nvec{v}}
\def\uu{\nvec{u}}

\begin{document}








\begin{center}{GENERALIZED RELATIVISTIC KINEMATICS}
\end{center}
\begin{center}S. N. Manida
\end{center}
\begin{center}{\it Saint-Petersburg State University, Saint-Petersburg, 198504, Russia, e-mail: sergey@manida.com}
\end{center}


\begin{abstract}
We propose a method for deforming an extended Galilei algebra that leads to a nonstandard realization
of the Poincar\'e group with the Fock-Lorentz linear fractional transformations. The invariant parameter
in these transformations has the dimension of length. Combining this deformation with the standard one
(with an invariant velocity $c$) leads to the algebra of the symmetry group of the anti-de Sitter space
in Beltrami coordinates. In this case, the action for free point particles contains the dimensional constants $R$ and $c$.
The limit transitions lead to the ordinary ($R\to \infty$) or alternative ($c\to \infty$) but nevertheless relativistic kinematics.

\end{abstract}




\renewcommand{\thefootnote}{\arabic{footnote}}
\setcounter{footnote}{0}

\section{Introduction}
The concept of inertial reference system underlies all classical and modern physics \cite{gal}.
The assumption of the existence of at least two such reference systems in an isotropic space leads
inevitably to the existence of an infinite set of such systems moving
with constant velocities with respect to one another\footnote[1]{To avoid confusion,
we call the velocity of a certain point
(the reference point for spatial distances) the velocity of reference system motion.}
The laws of nature must be formulated in terms that are covariant under transformations from one
inertial system to another. It is interesting to determine the general form of transformations between
arbitrary inertia reference systems, which also leads to determining the general form of spaces in which
such reference systems exist.

Since the first years of existence of the special theory of relativity, it has been known
 \cite{von1}-\cite{c1}, that the Lorentz transformation is the most general  {\it linear}
transformation of such type, and the Minkowski space is the corresponding space. An arbitrary constant $c$
having the dimension of velocity arises (is not postulated!) in deriving these transformations.
Fock showed \cite{fock}  that a linear fractional transformation is the most general form of such transformations.
He also showed that requiring that the transformations be linear is equivalent to postulating that a certain
velocity $c$ must be constant, which agrees with requiring that the Maxwell equations be invariant.
Based on the Fock papers, the explicit forms of such linear fractional transformations were constructed
in \cite{man},\cite{man2}  It turned out that these transformations contained not only
the constant $c$ but also a certain constant $R$ of the dimension of length; in this case, $c$
turned out to be invariant but not constant.
These results were confirmed from other standpoints by several authors
 \cite{step1} -\cite{mag}.
The local invariance of the Maxwell equations in this case was obtained automatically because the locally
linear fractional transformations coincided with ordinary Lorentz transformations.
The space in which the Fock-Lorentz linear fractional transformations are applicable differs from the Minkowski space,
although the algebra of symmetry generators of such a space coincides with that of the Poincar\'e group.
This space (the $R$ space in what follows)
can be obtained from the Minkowski space by the simple transformation
\be
  {\bf r}\Rightarrow \frac{{\bf r}R}{tc},
\hspace{20pt}
  {t}\Rightarrow -\frac{R^2}{tc^2}.  \label{rt}
\ee
In the case of such a transformation, the quantity ${\bf v}^2/c^2$, which is standard for many
equalities of the special theory of relativity becomes $({\bf r}-{\bf v}t)^2/R^2.$
One of the integrals of motion for $N$ free point particles in the special theory of relativity
$$
E=\sum_{n=1}^{N}\frac{m_nc^2  }{ \sqrt{1-{{\bf v}_n^2/ c^2}}}
$$
  becomes
\be
E\approx\sum_{n=1}^{N} m_nc^2+
\sum_{n=1}^{N}m_n\frac{{\bf v}_n^2}{ 2}.  \label{kin}
\ee
in the nonrelativistic limit  $v\ll c$.
A similar integral in the $R$ space has the form
$$
E=\sum_{n=1}^{N}\frac{m_nR^2  }{ \sqrt{1-{({\bf r}_n- {\bf v}_nt)^2/ R^2}}}
$$
and becomes
\be
E\approx \sum_{n=1}^{N}m_nR^2+\sum_{n=1}^{N}
m_n\frac{({\bf r}_n-
{\bf v}_nt)^2}{ 2}. \label{er}
\ee
in the ``noncosmological'' limit $|\vr-\vv t|\ll R$.
It seems that this expression is not related to the ordinary nonrelativistic kinematics.
Indeed, the conservation of the quantity
\be
\sum_{n=1}^{N}
m_n({\bf r}_n-
{\bf v}_nt)   \label{g}
\ee
in the nonrelativistic kinematics is obvious: it results from the symmetry under the group
of homogeneous Galilei transformations  ${\bf r}'={\bf r}+{\bf v}t,$ $t'=t$
and means that the velocity of motion of the center of mass of the system of material points is constant.
But the conservation of quantity $(\ref{g})$ does not imply that the sum of the squares in the right-hand side of
$(\ref{er})$ is preserved. Nevertheless, quantity $(\ref{er})$ is conserved in elastic collisions of point particles.
Because quantity $(\ref{kin})$ becomes $(\ref{er})$ under transformation $(\ref{rt}),$ the conservation
of quantity $(\ref{er})$ must be related to the symmetry transformation obtained from time shifts by transformation
$(\ref{rt}):$
\be
  t'=t+t_0;\quad \vr'=\vr \quad \Longrightarrow \quad
  t'=\frac{tt_0}{ t+t_0};\quad \vr'=\vr \frac{t_0}{ t+t_0}.   \label{in}
\ee
 The free-motion trajectory of a point particle from the initial point  ${\bf r}_0$ with the velocity ${\bf v}_0$
  has the form
   ${\bf r}(t ) = {\bf r}_0
+ {\bf v}_0t$.
This equation can be rewritten as  ${\bf r}/t = {\bf v}_0 + {\bf
r}_0/t$.
In such a representation,  $t$ and the coordinate ${\bf r}$ are replaced with $1/t$ and ${\bf r}/t,$
the initial values
of the coordinate and velocity are interchanged, but the trajectory remains straight. Consequently,
these new variables ${\bf r}/t$ and $1/t$ satisfy the same equations as the old ones ${\bf r}$ and $t$.
Such a symmetry of the nonrelativistic motion of free point particles indeed exists. It was shown in \cite{Sreed}
that the maximum continuous symmetry group for nonrelativistic inertial motion is 12-dimensional.
Here, we consider spaces in which the relativistic inertial motion has symmetries similar to (\ref{in}).
In Sec. 2, we determine the maximum symmetry group for nonrelativistic inertial motion, namely,
an extended Galilei group. This section is methodological. Here, we introduce notions required
for the further exposition.

In Sec.~3, we present two methods for deforming subalgebras of the algebra of the extended Galilei
group introduced in Sec.~2. These deformations lead to either the algebra of the Poincar\'e group and
the Minkowski space or to the algebra of the alternative (second) Poincar\'e group containing the
Fock--Lorentz linear fractional transformations and, consequently, to the R space. The Fock--Lorentz
linear fractional transformations are transformations in the coordinate space. In  \cite{45}, similar
transformations were applied to the momentum space with an invariant parameter of the dimension of
energy, which can be regarded as the Planck energy. But problems in interpreting the transformations
for the system of particles arise if this approach is used.
In Sec.~4, we present the combined deformation of the Galilei algebra leading to the algebra of the
symmetry group of the anti--de~Sitter space in Beltrami coordinates. Various versions of the relativistic
kinematics related to the de Sitter and anti--de~Sitter space in Beltrami coordinates were previously
considered in detail in \cite{41}--\cite{48}. But the relation between the symmetry of the anti-de Sitter space
and the Fock--Lorentz transformations and the corresponding symmetry group was not established previously.
We present a possible interpretation of our results in the conclusion.

\section{Maximum symmetry group of nonrelativistic inertial motion} \label{eg}

It is well known that the equations of motion for free nonrelativistic particles in a
$d$-dimensional space are invariant under the continuous group ${\cal G}_0$ of transformations
consisting of shifts of the time reference point $T$ and the static Galilei group
$$ G = SO(d) \ltimes ( P_d \otimes K_d )$$
the semidirect product of $SO(d)$  (the group of $d$--dimensional rotations without reflections) and the group
$P_d$  of $d$--dimensional translations and the group $K_d.$ of homogeneous direct Galilei transformations.
Because $T$ commutes with $SO(d),$ the complete group can be written as
$$ {\cal G}_0 = (T\otimes SO(d))\ltimes (P_d \otimes K_d ). $$

What is the maximum continuous invariance group for the motion
of nonrelativistic free point particles? The answer to this question
can be obtained by directly solving the Killing equations for the action
\be
S = \sum_{n=1}^N\frac{m_n}{ 2}\int \left(\frac{d{\bf r}_n}{ dt}\right)^2~dt.
\label{o}
\ee
We consider one-parameter infinitesimal transformations of coordinates and time:
\be
 t' = t + \varepsilon T({\bf r},t),\hspace{100pt} \label{d}
\ee
\be
  r'_i = r_i + \varepsilon X_i({\bf r},t),\, \, \,
\, \, \, \, \, \, \, \, \,
i = 1, \cdots, d.
\label{t}
\ee
 We obtain equations for the functions  $T({\bf r},t)$ and $X_i({\bf r},t)$
 from the invariance condition for action  (\ref{o}) up to a constant:
\be
\frac{1}{2}\left({\dot{\bf r}}\right)^2
 \left(
 {\partial_t T} + {\partial_j T }
 {\dot{r}_j}
 \right) -
 {\dot{r}_i}\left(
 {\partial_t X_i } + {\partial_j X_i }
 {\dot{r}_j} \right) =
 {\partial_t F }    +
 {\partial_i F}{\dot{r}_i},   \label{ch}
 \ee
where  $F({\bf r},t)$  is an arbitrary function of coordinates and time.
Here and hereafter, the summation over the repeated subscripts
 $i, j, k,\dots$ is assumed, and we use the notation
$$
\partial_i\equiv \frac{\partial }{ \partial r_i},\hspace{20pt}
\partial_t\equiv \frac{\partial }{ \partial t},\hspace{20pt}
\dot{f}\equiv \frac{df }{ dt}.
$$
Separating coefficients of the same powers of ${\dot{r}_i}$ in (\ref{ch})
we obtain the system of equations
 $$
 {\partial_t F } = 0; \hspace{5mm}
 {\partial_i T } = 0;   \hspace{5mm}
 {\partial_t X_i } =   {\partial_i F };
 \hspace{5mm}
 \frac{1}{ 2}\delta_{ij}{\partial_t T} =
 {\partial_j X_i }.
 $$
 The general solution of this system is
 \be
  F({\bf r}) = \frac{1}{ 2} c_1 {\bf r}^2 + c_4;
 \hspace{5mm}
 T(t) = 2c_2t - c_1t^2 + c_3;
 \hspace{5mm}
 {\bf X}({\bf r},t)= (c_2 - c_1t){\bf r},    \label{p}
 \ee
 where $c_1, c_2, c_3, c_4$ are arbitrary constants.

 If solution  (\ref{p})) is taken into account,
 then the generators of transformations (\ref{d}), (\ref{t}) become
 \bea
c_1\Longrightarrow
&
\hspace{20pt}
&  A= t r_i{\partial_i } + t^2{\partial_t }.
 \label{sh}
\\
c_2\Longrightarrow &
\hspace{10mm}
& M = - r_i{\partial_i } - 2 t
 {\partial_t };
\label{se}
\\
c_3\Longrightarrow &
\hspace{10mm}
&H = - {\partial_t};
\label{vo}
 \eea

 The commutation relations of generators (\ref{sh})--(\ref{vo})
 $$
   [M,H] = 2H; \hspace{15mm} [M,A] =- 2A; \hspace{15mm} [H,A] = M,
 $$
 indicate the  $SL(2,R)$ symmetry group.
 Indeed, the corresponding final transformations
 \bea
A  \Longrightarrow &
\hspace{30pt}
&t' = \frac{t }{ 1 + at};
\hspace{35pt}
 r'_i = \frac{r_i}{ 1 +
 at},
\label{de}
\\
M \Longrightarrow  &
\hspace{30pt}
&t' = (1+m)^2t;
\hspace{20pt}
r_i' = (1+m)r_i;
 \label{des}
\\
H \Longrightarrow   &
\hspace{30pt}
&t' = t+ h;
\hspace{40pt}
r_i' = r_i;
 \label{oo}
 \eea
 can be represented in the general form
 \be
 t'=\frac{\alpha t + \beta }{ \gamma t + \delta};
 \hspace{5mm}
 r'_i = \frac{r_i }{ \gamma t + \delta};
 \hspace{5mm}
 \alpha \delta - \beta \gamma = 1.  \label{od}
 \ee
Using the Noether theorem, we obtain the  conserved quantities
 \bea
 &
&{\cal A} = \frac{1}{2}
\sum_{n=1}^Nm_n(t\dot{\bf r}_n-{\bf r}_n)^2,
\label{9a} \\
&
&{\cal M} =
\sum_{n=1}^Nm_n(t\dot{\bf r}_n-{\bf r}_n)\dot{\bf r}_n,
 \label{10a}     \\
&
&{\cal H} = \frac{1}{2}
\sum_{n=1}^Nm_n{\dot{\bf r}_n}^2,
 \label{11a}
 \eea
corresponding to these transformations.

 Considering transformations similar to  (\ref{d}) and (\ref{t}) but with vector
  $\varepsilon_i$
 or tensor  $\varepsilon_{ij}$ parameters, we respectively obtain the group
 of spatial translations and homogeneous Galilei transformations
 or the group of rotations and represent the conserved quantities in the forms
 \be
{\bf P} =
\sum_{n=1}^N m_n\dot{\bf r}_n,
\label{9b}
\ee
\be
{\bf K} =
\sum_{n=1}^N m_n(t\dot{\bf r}_n-{\bf r}_n),
\label{10b}
\ee
\be
{\bf J_{ij}} =
\sum_{n=1}^N m_n\left((\dot{\bf r}_n)_i
(t\dot{\bf r}_n-{\bf r}_n)_j-
(\dot{\bf r}_n)_j(t\dot{\bf r}_n-{\bf r}_n)_i\right).
\label{11b}
 \ee
The nonstandard representation for angular momentum  $(\ref{11b})$
is introduced to stress that all conserved quantities
$(\ref{9a})$--$(\ref{11b})$  turn out to be linear or quadratic
functions of the kinematic quantities
$\dot{\bf r}_n$ and ${\bf r}_{0n}={\bf r}_n - t\dot{\bf r}_n.$

 The final result is as follows: the maximum invariance group for the nonrelativistic
 inertial motion of point particles is the group of the dimension  $(d^2+3d+6)/2$
 (the 12-dimensional group in the three-dimensional space)
$$
{\cal G} =
(SL(2,R)\otimes SO(d)) \ltimes (P_d \otimes K_d ).
$$

This result was first published in
 \cite{O'Raif}, \cite{Sreed}, although it was established as long as thirty years ago \cite{Sch}--\cite{Sch3}
that the group  ${\cal G}$ is the maximum symmetry group of solutions of the free
Schr\"odinger equation.

It is interesting that the invariance under transformation  (\ref{od})
was observed previously in various physical problems.
The closest example is the nonrelativistic
motion of a charged point particle in the field of an infinitely
heavy monopole  \cite{monopol}.

For the subsequent discussion, we note that the group  $SL(2,R)$
of transformations contains not only ordinary time shifts (\ref{oo}) with the generator
$H$ but also ``inverse'' time shifts (\ref{de}) produced by the generator  $A:$
\be
\frac{1}{ t'}=\frac{1}{ t}+a;
\hspace{10mm}
\frac{{\bf r}'}{ t'}=\frac{{\bf r}}{ t}.\label{AA}
\ee

In addition, we consider the one-parameter subgroup  $SO(2)$ of $SL(2,R),$
 generated by
\be
B\equiv \tau H-\frac{1}{ \tau}A,\label{B}
\ee
where we introduce an arbitrary constant   $\tau\ne 0.$
The final transformations corresponding to the generator
 $B$ have the forms
 \be
t'=\tau\frac{t\cos \alpha -
\tau\sin \alpha }{ t\sin \alpha + \tau\cos \alpha};
\hspace{10mm}
{\bf r}'=\tau\frac{{\bf r}}{ t\sin \alpha + \tau\cos \alpha}.   \label{ot}
\ee
This group contains finite cyclic subgroups  $C_n,$
consisting of powers of the elements  $\sigma_n$ --
rotations through the angles  $\alpha = 2\pi/n.$
Among them, the group   $C_4,$ consisting of powers of the ``inversion'' $\sigma_4$
(transformation (\ref{ot}) with $\alpha = \pi/2)$ is especially interesting:
\be
\sigma_4:
\hspace{40pt}
t'=-\frac{ \tau^2 }{ t};
\hspace{30pt}
{\bf r}'={\bf r}\frac{\tau}{t};
\hspace{35pt}
\label{och}
\ee
\be
(\sigma_4)^2:
\hspace{33pt}
t'=t;
\hspace{48pt}
{\bf r}'=-{\bf r};
\hspace{46pt}
\label{op}
\ee
\be
(\sigma_4)^3:
\hspace{33pt}
t'=-\frac{ \tau^2 }{ t};
\hspace{28pt}
{\bf r}'=-{\bf r}\frac{\tau}{ t}.
\hspace{35pt}
\label{osh}
\ee
 We note that  $(\sigma_4)^2=\sigma_2$  is the operation of reflection of
 spatial coordinates. Discrete operation  (\ref{op}) is usually added to the group  $SO(d),$
of rotations; the group  $O(d)$ is thus obtained.  Our analysis shows that the discrete transformation
(in the usual Galilei space-time) of "parity" (reflection of spatial coordinates) is contained in the
continuous group  $SL(2,R).$

We introduce the discrete operation $\tilde{\sigma},$ of time reversal:
\be
t'=-t, \hspace{10mm} {\bf r}'={\bf r}.
\label{ose}
\ee
It is hence clear that the symmetry transformations constructed
above are realized in the Mobius space and not in the usual
nonrelativistic space-time. The operation   $\tilde{\sigma}$
is not contained in the group  $SL(2,R),$ but is a symmetry of action  (\ref{o}).
The point is that the group  ${\cal G}$ was constructed above as the maximum
{\it continuous} invariance group of action  (\ref{o}).

We multiply an arbitrary element  $\Sigma\in SL(2,R)$
 by $\tilde{\sigma}:$
$$
  \widetilde{\Sigma}= \tilde{\sigma}\Sigma.
$$
Because $\tilde{\sigma}^2=I,$ the elements
$\widetilde{\Sigma}$ form the group
$\widetilde{SL}(2,R)=\Z_2\otimes SL(2,R)$
), and the complete maximum symmetry group has the form
$$
{\widetilde{\cal G}} = SO(d) \ltimes (\widetilde{SL}(2,R))
\ltimes (P_d \otimes K_d)).
$$

We note that integrals of motion  (\ref{9a})--(\ref{11b}) transform into
each other (up to the sign and dimensional factors,
powers of the parameter  $\tau$) or remain unchanged under the transformations $(\sigma_4)^n$ and
$\tilde{\sigma}(\sigma_4)^n.$

\bigskip

\section{Symmetry group of relativistic inertial motion}
\label{Poincare}

It is almost obvious that some of the above symmetries of inertial motion
are violated in the passage to the relativistic case. If the Killing
equation for the action
\be
S = -\sum_{n=1}^N m_n c^2\int \sqrt{1-\frac{\dot{\bf r}_n^2}{ c^2}}~dt, \label{ovo}
\ee
describing the motion of a relativistic free point particle is solved, then it
is easy to obtain the well-known result: the maximum invariance group of action  (\ref{ovo})
�is the Poincar\'e group, the semidirect product of the Lorentz group � $O(d,1)$
and the group of space-time translations
$P_d \otimes T.$
The discrete transformations of spatial coordinate reflection and time reversal are contained in the group $O(d,1).$
). The extension of the symmetry under time shifts up to  $SL(2,R),$
as in the case of the Galilei group, is not realized here.

We consider the ``relativization'' procedure for the Galilei group
${\cal G}_0$
up to the Poincar\'e group. We represent the generators of the group  ${\cal G}_0$ in the forms
\be
K_i = t {\partial_i};
\hspace{10mm}
P_i=  -{\partial_i };
\hspace{10mm}
H=-{\partial_t};
\hspace{10mm}
J_i=-\varepsilon_{ijk}{r_j\partial_k },
 \label{ode}
\hspace{10mm}
\ee
Here and hereafter, the space is three-dimensional  $(d=3),$
 and the tensor generators  ${J_{ij}}$ of the rotation group can therefore be represented in the form of the pseudovector
  ${J_i\equiv \varepsilon_{ijk} J_{jk}}.$
The algebra of these generators has the form
$$
\begin{array}{lll}
\, [J_i, J_j]=\ve_{ijk} J_k;&
[J_i, K_j]=\ve_{ijk} K_k;&
[J_i, P_j]=\ve_{ijk} P_k;\\
\, [J_i, H] = 0; &
 [K_i, H] = -P_i; &
 [P_i, H] = 0;  \\
\,
 [P_i, P_j] = 0; &
 [P_i, K_j] = 0;
&
[K_i, K_j] = 0.
\end{array}
$$
The passage to the Poincar\'e group is related to replacing the homogeneous
Galilei transformation with the homogeneous Lorentz transformation,
which is equivalent to replacing the generator  $K_i$ with the generator
\be
L_i = t \partial_i +
\frac{1}{ c^2}r_i \partial_t,  \label{dvo}
\ee
where we introduce the constant  $c$ with the dimension of velocity.
Just the algebra of generators  $\{J_i; L_i; P_i; H\}$ is
the algebra of the Poincar\'e group (the first Poincar\'e algebra below):
$$
\hspace{1pt}[J_i, J_j]=\ve_{ijk} J_k;
\hspace{20pt}[J_i, L_j]=\ve_{ijk} L_k;
\hspace{20pt}[J_i, P_j]=\ve_{ijk} P_k;
\hspace{29pt}
$$
$$
\hspace{2pt}[J_i, H] = 0;
\hspace{43pt}[L_i, H] = -P_i;
\hspace{31pt}[P_i, H] = 0;
\hspace{53pt}
$$
$$
\hspace{10pt}[P_i, P_j] = 0;
\hspace{40pt}[P_i, L_j] = \frac{1}{ c^2}\delta_{ij}H;
\hspace{17pt}[L_i, L_j] = -\frac{1}{ c^2}\ve_{ijk}J_k.
\hspace{13pt}
$$
Action  (\ref{ovo})  is invariant under the group of transformations produced by
the generators  $\{J_i; L_i; P_i; H\},$
and the discrete transformations of time reversal and spatial coordinate reflection.

We note that the complete invariance group for nonrelativistic motion ${\cal G}$
 contains not only the subgroup ${\cal G}_0,$
 but also a similar subgroup ${\cal G}_1,$ produced by the generators    $\{J_i; P_i; K_i; A\}$
with the algebra
$$
\hspace{10pt}[J_i, J_j]=\ve_{ijk} J_k;
\hspace{40pt}[J_i, K_j]=\ve_{ijk} K_k;
\hspace{17pt}[J_i, P_j]=\ve_{ijk} P_k;
\hspace{13pt}
$$
$$
\hspace{-32pt}[J_i, A] = 0;
\hspace{65pt}[K_i, A] = 0;
\hspace{45pt}[P_i, A] = -K_i;
\hspace{-21pt}
$$
$$
\hspace{-22pt}[P_i, P_j] = 0;
\hspace{62pt}[P_i, K_j] = 0;
\hspace{41pt}[K_i, K_j] = 0.
\hspace{-1pt}
$$

This algebra can be deformed to obtain the algebra of the Poincar\'e group by replacing the translations
$P_i$ with the ``inverted'' translations
\be
F_i=  -{\partial_i }+\frac{1}{ R^2}(tr_i\partial_t+r_ir_k\partial_k),
\label{dv1}
\ee
where $R$ is a constant with the dimension of length. The term ``inverted'' is used because generator
(\ref{dv1}) is obtained from the generator  $L_i$  under the transformation
$\sigma_4$ or $\tilde{\sigma}_4=\tilde{\sigma}\sigma_4$ with the parameter
$\tau=R/c,$  just as all generators of the group ${\cal G}_1$ are obtained from the generators of the group  ${\cal G}_0.$

The deformed algebra
$$
\hspace{3pt}[J_i, J_j]=\ve_{ijk} J_k;
\hspace{20pt}[J_i, K_j]=\ve_{ijk} K_k;
\hspace{20pt}[J_i, F_j]=\ve_{ijk} F_k;
\hspace{27pt}
$$
$$
\hspace{1pt}[J_i, A] = 0;
\hspace{45pt}[F_i, A] = -K_i;
\hspace{35pt}[K_i, A] = 0;
\hspace{48pt}
$$
$$
\hspace{13pt}[K_i, K_j] = 0;
\hspace{37pt}[K_i, F_j] = \frac{1}{ R^2}\delta_{ij}A;
\hspace{15pt}[F_i, F_j] = -\frac{1}{ R^2}\ve_{ijk}J_k
\hspace{16pt}
$$
forms the second representation of the algebra of the Poincar\'e group (the second Poincar\'e algebra
below)\footnote[2]{A similar concept of "second Poincar\'e group" was introduced in (\cite{20});
its algebra of generators was obtained as a result of contracting the de Sitter algebra a $R\to 0$.}

The final transformations produced by the generators  $F_i$
\be
t'=\frac{t\sqrt{1-{a^2}/{R^2}}}
{1-{({\bf r}{\bf a})}/{R^2}},
\hspace{20pt}
{\bf r}'_{||}=
\frac{{\bf r}_{||}-{\bf a}}
{1-{({\bf r}{\bf a})}/{R^2}},
\hspace{20pt}
{\bf r}'_{\perp}=
\frac{{\bf r}_{\perp}\sqrt{1-{a^2}/{R^2}}}
{1-{({\bf r}{\bf a})}/{R^2}},
\label{zz}
\ee
where we introduce the notation
$$
{\bf r}_{||}={\bf n}{({\bf r}{\bf n})},
\hspace{30pt}
{\bf r}_{\perp}={\bf r}  -{\bf r}_{||},
\hspace{30pt}
{\bf n}=\frac{\bf a}{|\bf{a}|},
$$
constitute the Lorentz transformations for the quantities
$$
\tilde{\bf r}=\frac{\bf r}{t};
\hspace{30pt}
\tilde{t}=\frac{R}{t}.
$$
Indeed, it follows from formulas (\ref{zz})  that
\be
\tilde{t}'=\gamma_a
\left(\tilde{t}-{\frac{({\bf a}\tilde{\bf r})}{ R}}\right),
\hspace{20pt}
\tilde{\bf r}'_{||}=\gamma_a
\left(\tilde{\bf r}_{||}-
{\bf a}\tilde{t}\right),
\hspace{20pt}
\tilde{\bf r}'_{\perp}=
\tilde{\bf r}_{\perp},
 \label{izz}
\ee
where $\gamma_a^{-1}=\sqrt{1-\frac{a^2}{ R^2}}.$

To construct the action that is invariant under the transformations produced by the generators
  $\{J_i; F_i; K_i; A\},$ we take into account that ordinary relativistic action (\ref{ovo}) is
\be
S = -\sum_{n=1}^Nm_nc\int \sqrt{ds_n^2},   \label{18a}
\ee
where
\be
ds_n^2=c^2dt^2-d{\bf r}_n^2.   \label{ds}
\ee
The transformation   $\sigma_4$ translates linear element  (\ref{ds}) �
into that of the $R$ space introduced in
\cite{man},  \cite{man2}
\be
d\tilde{s}^2=\frac{\tau^2}{ t^4}[R^2dt^2-(td{\bf r}-{\bf r}dt)^2],\qquad R\equiv c\tau.
\label{ds2}
\ee
Substituting (\ref{ds2}) in (\ref{18a}), we obtain the action
\be
S = -\sum_{n=1}^Nm_nR^2\int \sqrt{1-\frac{(\dot{\bf r}_nt-
{\bf r}_n)^2}{ R^2}}~\frac{dt}{ t^2}.
\label{slf}
\ee
The invariance of action (\ref{slf}) under the transformations produced by the generators
{$\{J_i ; F_i; K_i;A \},$ } implies the conservation of the quantities
\be
A
\hspace{10pt}
\Rightarrow
\hspace{40pt}
{\cal H}=\sum_{n=1}^N\frac{m_n}{\sqrt{1-(\dot{\bf r}_nt-
{\bf r}_n)^2/R^2}};
\hspace{70pt}
\label{E}
\ee
\be
{K}_i
\hspace{10pt}
\Rightarrow
\hspace{40pt}
{\bf K}= \sum_{n=1}^N\frac{m_n(\dot{\bf r}_nt-{\bf r}_n)}{\sqrt{1-
(\dot{\bf r}_nt-{\bf r}_n)^2/R^2}};
\hspace{70pt}
\label{K}
\ee
\be
F_i
\hspace{10pt}
\Rightarrow
\hspace{40pt}
{\bf P}=
\sum_{n=1}^N\frac{m_n\dot{\bf r}_n}{\sqrt{1-
(\dot{\bf r}_nt-{\bf r}_n)^2/R^2}};
\hspace{70pt}
\label{F}
\ee
\be
J_i
\hspace{10pt}
\Rightarrow
\hspace{40pt}
{\bf J}=
\sum_{n=1}^N\frac{m_n({\bf r}_n-t\dot{\bf r}_n)\times\dot{\bf r}_n}{\sqrt{1-
(\dot{\bf r}_nt-{\bf r}_n)^2/R^2}}.
\hspace{70pt}
\label{J}
\ee

The dispersion relation follows from (\ref{E}) and  (\ref{K}):
\be
{\cal H}^2-\frac{{\bf K}^2}{ R^2}=m^2,     \label{onsh}
\ee
It can be seen from this relation that in the $R$ space, ${\cal H}$ is the energy and
 ${\bf K}$  is the momentum of the system of noninteracting point particles.
 We note that the free-particle energy is minimum if the relation between
 its velocity $v$ and the distance to the observer $r$ is $v=r/t.$
  In what follows, such velocities are called the ``Hubble'' velocities.
The velocity of light in this space, i.e., the velocity of a massless particle,
depends on the coordinates, the time, and the propagation direction  ${\bf n}$ $({\bf n}^2=1)$
\cite{man}:
\be
{\bf c}({\bf r}, t, {\bf n})=
{\bf n}\frac{R}{ t}
\sqrt{1-\frac{r^2}{ R^2}+\frac{({\bf r}{\bf n})^2}{ R^2}}+{\bf n}\frac{({\bf r}{\bf n})}{ t}.
\label{ct}
\ee
For the magnitude of the velocity of light measured in laboratories (i.e., for $r=0$), we obtain the expression
\be
{\bf c}(0, t, {\bf n})=
{\bf n}\frac{R}{ t}.
\label{ct0}
\ee
Despite the explicit ``dependence'' of the velocity of light on the time, distance,
and direction expressed by formulas
(\ref{ct}), (\ref{ct0})), a particular light pulse propagates along the light cone with a constant velocity.
Neither astrophysical observations of the spectra of remote sources nor optical observations
 in ground-based laboratories can, in principle, fix the time dependence of the velocity of light of form (\ref{ct0}),
  because at the instant $t_0$ at the point  $r=0$
 the observer detects the light from the source at the distance $r$ at a certain instant  $t<t_0.$ We obtain
 $c(r,t)=(R-r)/t$ from equality (\ref{ct}) and $c(0,t_0)=R/t_0$
from (\ref{ct0}).  These velocities coincide at $r(t)=R-tR/t_0,$
, which means that the light pulse propagates with a constant velocity $c_0=R/t_0.$
Moreover, we must emphasize that this velocity is the same for all observers at a given point in space-time,
which constitutes the fundamental postulate of the special theory of relativity. We note that equalities
(\ref{slf})--(\ref{onsh}) contain only one universal constant  $R,$ with the length dimension.
There is no constant with the dimension of velocity in these relations,
but they fit completely into our concept of relativistic theory.

The space described by formulas  (\ref{ds2})--(\ref{ct}),
 is symmetric under the Galilei transformations, linear fractional space translations (\ref{zz})
and time reversal translations
  (\ref{AA}). If these transformations are combined, then we can construct the transformation of the coordinates between two reference systems moving
  with a velocity ${\bf u}$ with respect to each other such
  that the coordinate origins coincide at the instant
   $t_0.$
 Just this coordinate transformation is the Fock-Lorentz linear fractional transformation
\cite{man}, \cite{man2}:
\bea
t'=
\frac
{t}
{\ga_0-\left(\ga_0-1\right)t/t_0-\ga_0
{\vr\uu t_0 }/{R^2}},
\label{F37a}\\
\vr'_{||}=
\frac
{\ga_0\left(\vr_{||}-\uu (t-t_0)\right)}
{\ga_0-\left(\ga_0-1\right)t/t_0-\ga_0
{\vr\uu t_0 /{R^2}}},
 \label{F37b} \\
\vr'_{\perp}=
\frac
{\vr_{\perp}}
{\ga_0-\left(\ga_0-1\right)t/t_0-\ga_0
{\vr\uu t_0}/{R^2}}.
\label{F37c}
 \eea
$$\ga_0\equiv 1/\sqrt{1-u^2t_{0}^2/R^2}, \,\,\,\,\,\,\,\,u<R/t_0$$

\section{Combined deformation of the Galilei algebra} \label{comb}
The abovementioned two deformations of the algebra of the Galilei group can be combined.
We consider the algebra of the generators
$\{J_i; F_i; L_i; B=({R}/{c})H-({c}/{R})A\}:$
$$ \hspace{0pt}[J_i, J_j]=\ve_{ijk} J_k;
\hspace{30pt}[J_i,
L_j]=\ve_{ijk} L_k;
\hspace{16pt}[J_i, F_j]=\ve_{ijk} F_k;
\hspace{61pt} $$ $$
\hspace{0pt}[J_i, B] = 0;
\hspace{55pt}[F_i, B] = \frac{c}{ R}L_i;
\hspace{25pt}[L_i, B] = -\frac{R}{ c}F_i;
\hspace{60pt}
$$
$$
\hspace{0pt}[L_i, L_j] = -\frac{1}{ c^2}\ve_{ijk}J_k;
\hspace{9pt}[F_i, L_j] = \frac{1}{ Rc}\delta_{ij}B;
\hspace{5pt}[F_i, F_j] = -\frac{1}{ R^2}\ve_{ijk}J_k.
\hspace{38pt}
$$
The In\"on\"u-Wigner contraction  \cite{IW}, \cite{IW2} of the constructed algebra is obvious.
We obtain the first Poincar\'e algebra in the limit
 $R\rightarrow \infty$ and the second in the limit  $c\rightarrow \infty.$

The presence of two dimensional constants  $R,$ $c$
allows making all generators and structure algebraic constants dimensionless.
For this, we introduce the dimensionless representations of the generators
$$
\hat{L}_i=cL_i= ct \partial_i +
\frac{1}{ c}r_i \partial_t,
$$
$$
\hat{F}_i=RF_i=
-R{\partial_i }+\frac{1}{ R}(tr_i\partial_t+r_ir_k\partial_k).
$$
In this representation, the commutation relations become
$$
\hspace{4pt}[J_i, J_j]=\ve_{ijk} J_k;
\hspace{22pt}[J_i, \hat{L}_j]=\ve_{ijk} \hat{L}_k;
\hspace{20pt}[J_i, \hat{F}_j]=\ve_{ijk} \hat{F}_k;
\hspace{24pt}
$$
$$
\hspace{6pt}[J_i, B] = 0;
\hspace{46pt}[\hat{F}_i, B] = \hat{L}_i;
\hspace{41pt}[\hat{L}_i, B] = -\hat{F}_i;
\hspace{36pt}
$$
$$
\hspace{17pt}[\hat{L}_i, \hat{L}_j] = -\ve_{ijk}J_k;
\hspace{10pt}[\hat{F}_i, \hat{L}_j] = \delta_{ij}B;
\hspace{27pt}[\hat{F}_i, \hat{F}_j] = -\ve_{ijk}J_k;
\hspace{27pt}
$$
and constitute the algebra  $AdS(3,2)$ of the anti-de Sitter group.

The final transformations of the space-time produced by the generators
      $\hat{L}_i$ are ordinary Lorentz transformations, those produced by the generators
$\hat{F}_i$ are equalities (\ref{zz}), and those produced by the generators
 $B$ are equalities  (\ref{ot}).

We introduce the standard notation for the Lorentz vectors and their products:
$$
x^{\mu}=(ct, {\bf r}),
\hspace{20pt}
a^{\mu}=(a_0, {\bf a}),
\hspace{20pt}
  (ax)=a_0ct-{\bf a}{\bf r},
\hspace{20pt}
a^2=a_0^2-{\bf a}^2.
$$
In this notation, an arbitrary combination of transformations
(\ref{ot}) and (\ref{zz}) can be represented in the form
 \be
x'^{\mu}=\frac{x^{\mu}-a^{\mu}+(\gamma-1)a^{\mu}[(ax)/a^2-1]}
{\gamma(1+(ax)/R^2)},
\hspace{20pt}
\gamma^{-1}=\sqrt{1-a^2/R^2}.
\label{BF}
\ee
The group property of transformation  (\ref{BF})
can be verified by direct calculations. Moreover, transformation
  (\ref{BF}) coincides with  (\ref{zz}) for $a_0=0$ and with (\ref{B})
  for
${\bf a}=0,$ $a_0=c\tau \tg \alpha.$

Squaring equality  (\ref{BF}) we obtain
\be
x'^2=\frac{(x-a)^2+[a^2x^2-(ax)^2]/R^2}{(1+(ax)^2/R^2)^2}.   \label{sq}
\ee
Setting  $a^{\mu}=x^{\mu}+dx^{\mu}$ in(\ref{sq}),
we obtain a quantity that is invariant under the entire group of transformations
$AdS(d,2):$
$$
ds^2=R^2\frac{dx^2(R^2+x^2)-(xdx)^2}{(R^2+x^2)^2}
$$
and in the three-dimensional notation
\be
ds^2=\frac{R^4c^2dt^2}{(R^2+c^2t^2-{\bf r}^2)^2}
\left(
1-\frac{\dot{\bf r}^2}{c^2}-
\frac{(\dot{\bf r}t-{\bf r})^2}{R^2}
+\frac{({\bf r}\times\dot{\bf r})^2}{R^2c^2}
\right).\label{3ds}
\ee
The action for free point particles becomes
\be
 S=  -\sum_{n=1}^N\int\frac{m_nc^2R^2dt}{(R^2+c^2t^2-{\bf r}_n^2)}
\sqrt{1-\frac{\dot{\bf r}_n^2}{c^2}-
\frac{(\dot{\bf r}_nt-{\bf r}_n)^2}{R^2}
+\frac{({\bf r}_n\times\dot{\bf r}_n)^2}{R^2c^2}}. \label{sads}
\ee
In the limit  $c\to \infty$ action (\ref{sads}) transforms into action
(\ref{ds}).

The conserved quantities are similar to  (\ref{E})--(\ref{J}):
\be
\hspace{40pt}
{\cal H}=\sum_{n=1}^N\frac{m_n}{
\sqrt{1-{\dot{\bf r}_n^2}/{c^2}-
{(\dot{\bf r}_nt-{\bf r}_n)^2}/{R^2}
+{({\bf r}_n\times\dot{\bf r}_n)^2}/({Rc)^2}}},
\hspace{50pt}\label{Ea}
\ee
\be
\hspace{40pt}
{\bf K}=
\sum_{n=1}^N\frac{m_n(\dot{\bf r}_nt-{\bf r}_n)}{
\sqrt{1-{\dot{\bf r}_n^2}/{c^2}-
{(\dot{\bf r}_nt-{\bf r}_n)^2}/{R^2}
+{({\bf r}_n\times\dot{\bf r}_n)^2}/({Rc)^2}}},
\hspace{50pt}\label{Ka}
\ee
\be
\hspace{40pt}
{\bf P}=
\sum_{n=1}^N\frac{m_n\dot{\bf r}_n}{
\sqrt{1-{\dot{\bf r}_n^2}/{c^2}-
{(\dot{\bf r}_nt-{\bf r}_n)^2}/{R^2}
+{({\bf r}_n\times\dot{\bf r}_n)^2}/({Rc)^2}}},
\hspace{50pt}\label{Fa}
\ee
\be
\hspace{40pt}
{\bf J}=
\sum_{n=1}^N\frac{m_n[{\bf r}_n\times\dot{\bf r}_n]}{
\sqrt{1-{\dot{\bf r}_n^2}/{c^2}-
{(\dot{\bf r}_nt-{\bf r}_n)^2}/{R^2}
+{({\bf r}_n\times\dot{\bf r}_n)^2}/({Rc)^2}}}.
\hspace{50pt}\label{Ja}
\ee
Equalities (\ref{Ea})-(\ref{Ja}) imply the dispersion relation
\be
  {\cal H}^2-\frac{{\bf P}^2}{ c^2}-\frac{{\bf K}^2}{ R^2}+\frac{{\bf J}^2}{ c^2R^2}=m^2,  \label{dis}
\ee
which transforms into  (\ref{onsh}) in the limit  $c\to \infty$ and into
the ordinary relation of the traditional theory of relativity in the limit  $R\to \infty.$

The equation for light cones can be obtained from  (\ref{sads}). In the simple case  $\vr=0$
we obtain the speed of light  $c(t)$ from the relation
\be
\frac{1}{c(t)^2}=\frac{1}{c^2}+\frac{t^2}{R^2}. \label{cads}
\ee

The limit transition  $c\to \infty$  in (\ref{3ds})--(\ref{cads})
leads directly to  (\ref{ds2})--(\ref{onsh}).

\section{Conclusions}
At first glance, it seems that the space-time described by relations
 (\ref{ds2})--(\ref{ct}) is not related to the surrounding world.
 But if we consider spatial regions in a small neighborhood
 $|{\bf r}|\ll R,$  of the coordinate origin, time intervals $\Delta t$ in a small neighborhood of a point
   $t_0$
($\Delta t \ll t_0$) and velocities
$|\dot{\bf r}|\gg |{\bf r}|/t_0,$
then equalities  (\ref{ds2})--(\ref{ct})
transform into the ordinary relations of relativistic physics with the speed of light  $c_0\equiv R/t_0.$
�Consequently, the $R$ space can be regarded as a ``cosmological'' generalization of the Minkowski
space if $R$ is set equal to the radius of the visible part of the Universe  ($\sim 10^{26}$~�)
and $t_0$ is set to be the time passed after the Big Bang  ($\sim 10^{10}$~years or  $3\cdot 10^{18}$~�).
In this case, all relativistic formulas of the standard theory of relativity are applicable,
and the corrections to them have the cosmological character: their orders are  $r/R,$ $\Delta t/t_0,$
and
$|{\bf r}|/(t_0|\dot{\bf r}|).$  In the last expression, the quantity
   $|{\bf r}|/t_0$ is the ``Hubble'' velocity of the point  ${\bf r}$.
In this case, all relations of the standard relativistic physics hold if the relativistic body velocities considerably exceed the Hubble velocity.
For obviousness, we note that  even at distances of the order of the solar system dimensions ($\sim 10^{12}$~m),
the Hubble velocity is a very small quantity $3\cdot10^{-6}$~m/s), for which relativistic effects are not observed.

If the physics described by formulas (\ref{ds2})--(\ref{ct}),
of the $R$ space is really related to our world, then we must understand what the limit transition  $c\to \infty,$
translating the physics of the anti-de Sitter space into that of the $R$ space means. In this case, there is the constant
$R$ and the very large constant $c$ (compared with the observed speed of light $c_0$) in our world, and
the ratio $R/c$ is a small (compared with the Universe's lifetime) but fundamental constant,
which can be equal to the Planck time
\be
t_{PL}=\sqrt{\frac{G\hbar}{c_0^5}},\label{tau}
\ee
which is  $10^{60}$ times less than the Universe's lifetime. In this case, the speed of light at a given instant is also
 $10^{60}$ times less than the speed of light at the initial instant of the Universe's existence. In this way,
 interesting prospects are offered to correct our concepts of the first instants of the Universe's development and to
 introduce quantum gravitational parameters into physics at the level of kinematics and not dynamics. Moreover,
 the speed  $c_0$ in expression  (\ref{tau}) --
 is the physical speed of light (time dependent in the geometry under consideration).
 It hence follows that the constants  $\hbar,$  $G$ can also be time dependent. Hence, the Planck length
  $l_{PL}=c_0t_{PL}$ for example, is time dependent in this geometry, just as $c_0$ is.
  Consequently, the Planck length could be    $10^{60}$ times greater in the first instants of the Universe's existence,
   i.e., it could coincide with the dimension $R$ of the entire Universe.

{\bf Acknowledgments.}  A major part of this work would not have been done without
the constant communications, hot discussions,
and constructive criticism of A.~N.~Vasiliev over the course of many years.


\begin{thebibliography}{99}
\bibitem{gal}
1.  G. Galilei, "Dialogue about the two major world systems -
Ptolemeic and Copernican [in Russian]," in: Selected Works,
Vol. 1, Nauka, Moscow (1964), pp. 109-586.
\bibitem{von1}
W. von Ignatowsky,
{\it Phys. Z.}
{\bf 20} (1910) 972
\bibitem{von2}
W. von Ignatowsky, {\it Arch. Math. Phys. (Leipzig)} {\bf 17} (1910) 1
\bibitem{c1}
P. Frank and H. Rothe, {\it Ann. Phys. Lpz.} {\bf 34} (1911) 825
\bibitem{fock}
 V. Fock, The Theory of Space, Time, and Gravitation [in Russian],
Fizmatgiz, Moscow (1961); English transl., Pergamon, Oxford (1963).
\bibitem{man}
S. N. Manida,
{\it Fock-Lorentz transformations and time-varying speed of light,}
arXiv:gr-qc/9905046.
\bibitem{man2}
S. N. Manida, Vestnik St.-Petersburg. Gos. Univ. Ser. 4 : Fizika, khimiya, 2, 3-17 (2001).
\bibitem{step1} S. S. Stepanov, {\it Fundamental physical constants and the principle of parametric incompleteness},
arXiv:physics/9909009
\bibitem{step2} S. S. Stepanov,
{\it Phys. Rev.} D62 (2000) 023507;
arXiv:astro-ph/9909311.
\bibitem{mag} J. Magueijo,
{\it Phys. Rev.} D62 (2000) 103521, arXiv:gr-qc/0007036
\bibitem{Sreed}
O. Jahn, V.V. Sreedhar,
{\it Am.J.Phys.,} {\bf 69}, (2001) 1039-1043,
arXiv:math-phys/0102011
\bibitem{O'Raif}
L. O'Raifeartaig, V. V. Sreedhar,
{\it Annals Phys.,} {\bf 293} (2001) 215-227,
arXiv:hep-th/0007199
\bibitem{monopol} R. Jackiw,
{\it Ann. Phys.,} {\bf 129} (1980) 183-200
\bibitem{Sch} U. Niederer,
{\it Helvetica Physica Acta,} {\bf 45} (1972) 802;
\bibitem{Sch2}C. R. Hagen,
{\it Phys. Rev.,} {\bf D5} (1972) 377-388;
\bibitem{Sch3}R. Jackiw,
{\it Phys. Today} {\bf 25}
(1972) 23.
\bibitem{Weyl} H. Weyl,
{\it Raum, Zeit, Materie.} Berlin, 1923
\bibitem{45} J. Magueijo, L. Smolin, Phys. Rev. Lett., 88 (2002) 190403, arXiv:hep-th/0112090
\bibitem{41}
H.-Y. Guo, {\it Science in China A,} {\bf 51}:4 (2008) 568-603.
\bibitem{42}
H.-Y. Guo, C.-G. Huang, Y. Tian, Z. Xu, B. Zhou,
{\it A Model on cosmological constant as origin of inertia,}   arXiv:hep-th/0405137
\bibitem{43}
H.-Y. Guo, C.-G. Huang,  H.-T. Wu, B. Zhou,
{\it The principle of relativity, kinematics and algebraic relations,}   arXiv:0812.0871
\bibitem{44}
H.-Y. Guo, H.-T. Wu, B. Zhou,
{\it Phys. Lett. B,} {\bf} 670 (2009) 437-441,   arXiv:0809.3562
\bibitem{46}
 H.-Y. Guo, C.-G. Huang, Y. Tian, Z. Xu, B. Zhou,
{\it Class.Quant.Grav.,}{\bf 24} (2007) 4009-4036,
arXiv:gr-qc/0703078
\bibitem{47}
H.-Y. Guo, C.-G. Huang, Y. Tian, Z. Xu, B. Zhou, {\it Front.Phys.China}, {\bf 2} (2007) 358-363,
arXiv: hep-th/0607016
\bibitem{48}
H.-Y. Guo, {\it The Beltrami Model of De Sitter Space: From Snyder's quantized space-time to de Sitter invariant relativity},
arXiv: hep-th/0607017
\bibitem{20}
R. Aldrovandi and J. G. Pereira,
{\it A Second Poincar\'e Group,}   arXiv:gr-qc/9809061
\bibitem{IW}
E.~In\"on\"u, E.~P.~Wigner
{\it  Proc. Natl. Acad. Sci., USA} {\bf 39} (1953) 510
\bibitem{IW2}
E. In\"on\"u, E. P.  Wigner
{\it  Proc. Natl. Acad. Sci., USA} {\bf 40} (1954) 119









\end{thebibliography}
\end{document}